\documentclass[aps,prl,secnumarabic,nobibnotes,twocolumn,superscriptaddress,longbibliography]{revtex4-2}
\usepackage{amsfonts}
\usepackage{mathrsfs}
\usepackage{amsmath}
\usepackage{color}
\usepackage{natbib}
\usepackage{graphicx}
\usepackage{bm}
\usepackage{amssymb}
\usepackage{xspace}
\usepackage{epstopdf}
\usepackage{dcolumn}
\usepackage{multirow}
\usepackage[colorlinks=true, letterpaper=true, pdfstartview=FitV, linkcolor=blue, citecolor=blue, urlcolor=blue]{hyperref}
\usepackage{wrapfig}

\makeatletter

\newcommand{\Rmnum}[1]{\expandafter\@slowromancap\romannumeral #1@}
\makeatother

\begin{document}
\title{Unconventional topological Weyl-dipole  phonon}

\author{Jianhua Wang}
\address{Institute for Superconducting and Electronic Materials, Faculty of Engineering and Information Sciences, University of Wollongong, Wollongong 2500, Australia.}
\address{Institute of Quantum Materials and Devices, Tiangong University, Tianjin 300387, China.}

\author{Yang Wang}
\address{Key Lab of Advanced Optoelectronic Quantum Architecture and Measurement (MOE), Beijing Key Lab of Nanophotonics $\&$ Ultrafine Optoelectronic Systems, and School of Physics, Beijing Institute of Technology, Beijing 100081, China.}

\author{Feng Zhou}
\address{Institute of Quantum Materials and Devices, Tiangong University, Tianjin 300387, China.}

\author{Wenhong Wang}
\address{Institute of Quantum Materials and Devices, Tiangong University, Tianjin 300387, China.}

\author{Zhenxiang Cheng}
\address{Institute for Superconducting and Electronic Materials, Faculty of Engineering and Information Sciences, University of Wollongong, Wollongong 2500, Australia.}

\author{Shifeng Qian}\thanks{Corresponding author}\email{qiansf@ahnu.edu.cn}
\address{Anhui Province Key Laboratory for Control and Applications of Optoelectronic Information Materials, Department of Physics, Anhui Normal University, Wuhu, Anhui 241000, China.}

\author{Xiaotian Wang}\thanks{Corresponding author}\email{xiaotianw@uow.edu.au}
\address{Institute for Superconducting and Electronic Materials, Faculty of Engineering and Information Sciences, University of Wollongong, Wollongong 2500, Australia.}

\author{Zhi-Ming Yu}
\address{Key Lab of Advanced Optoelectronic Quantum Architecture and Measurement (MOE), Beijing Key Lab of Nanophotonics $\&$ Ultrafine Optoelectronic Systems, and School of Physics, Beijing Institute of Technology, Beijing 100081, China.}

\begin{abstract}
A pair of Weyl points (WPs) with opposite Chern numbers  ${\cal{C}}$ can exhibit an additional higher-order $Z_2$ topological charge, giving rise to the formation of a $Z_2$ Weyl dipole.
Owing to the nontrivial topological  charge, $Z_2$ Weyl dipoles should also appear in pairs, and  the WPs within each $Z_2$ Weyl dipole can not be annihilated when meeting together.
As a novel topological state, the topological Weyl-dipole phase (TWDP) has garnered significant attention, yet its realization in crystalline materials remains a challenge.
Here, through first-principles calculations and theoretical analysis, we demonstrate the existence of the  Weyl-dipole phase  in the phonon spectra of the $P6_3$ type Y(OH)$_3$.
Particularly, the Weyl dipole in this system is protected  by  a quantized quadrupole moment, and it distinguished from conventional Weyl dipole, as it comprises an unconventional charge-3  WP with ${\cal{C}}=-3$  and three conventional charge-1  WPs with ${\cal{C}}=1$.
Consequently,  the Weyl-dipole phase  in Y(OH)$_3$ features unique two-dimensional (2D) sextuple-helicoid  Fermi-arc states on the top and bottom  surfaces, protected by the Chern number, as well as one-dimensional (1D) hinge states that connect the two Weyl dipoles along the side hinges, guaranteed by the quantized quadrupole moment.
Our findings not only introduce a novel higher-order  topological phase, but also promote  Y(OH)$_3$ as  a promising platform for exploring multi-dimensional boundaries and the interaction between first-order and second-order topologies.
\end{abstract}
\maketitle

\begin{figure}
\includegraphics[width=8.1cm]{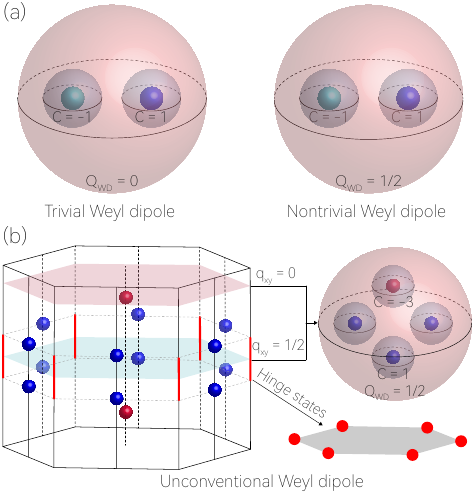}
\caption{(a) Schematic illustration of trivial and nontrivial Weyl dipoles. (b) Schematic illustration of the newly proposed unconventional Weyl dipole, consisting of Weyl complex, i.e., two  charge-3 (C-3) WPs with Chern number of -3 and six  charge-1 (C-1) WPs with Chern number of +1, characterized by a quantized quadrupole moment $Q_{WD}$ = 1/2.
\label{fig1}}
\end{figure}

\textcolor{blue}{\textit{Introduction.--}}
The investigation into topological quantum states enhances our understanding of electronic behavior in solid materials~\cite{RevModPhys.82.3045,RN316,ref3,RN317,YU2022375}.
Thus far, various topological  band crossings have been proposed~\cite{YU2022375}, including  nodal points~\cite{PhysRevLett.121.035302,PhysRevLett.120.016401,PhysRevLett.123.065501,PhysRevLett.122.104302,PhysRevLett.124.105303,PhysRevLett.126.185301,PhysRevLett.131.116602,
https://doi.org/10.1002/advs.202207508,RN216} and high-dimensional geometric shapes~\cite{PhysRevLett.123.245302,RN108,RN106,RN318,10.1063/5.0095281,doi:10.1126/science.adf8458}.
Notably, the WPs, exhibiting zero-dimensional (0D) twofold degeneracy in three-dimensional (3D) momentum space and carrying topological  $Z$ charge defined by the Chern number, have been regarded as one of the most important  topological degeneracies.
WPs usually appear in pairs with opposite Chern numbers. The schematic of the classic Weyl pair, composed of two conventional WPs characterized by Chern numbers of $\pm 1$, is shown in Fig.~\ref{fig1}(a).
Due to the bulk-boundary correspondence, topological Weyl phases should have  a  topologically protected  2D Fermi-arc state  on the surface of the system.

The recent discovery of higher-order topological phases~\cite{PhysRevLett.120.026801,PhysRevLett.121.106403,PhysRevB.98.241103,PhysRevLett.123.216803,PhysRevLett.123.256402,PhysRevB.100.205406,
PhysRevLett.125.146401,PhysRevLett.125.266804,RN319,RN320,PhysRevB.104.085205,PhysRevB.103.205123,RN321,
RN322,https://doi.org/10.1002/advs.202202564,PhysRevLett.128.026405,PhysRevLett.130.116103,PhysRevB.107.245116,https://doi.org/10.1002/advs.202301952,
https://doi.org/10.1002/adfm.202304499,https://doi.org/10.1002/adma.202402232,Yang_2024,https://doi.org/10.1002/adfm.202316079,PhysRevB.110.115427}, which host boundary states in at least two dimensions lower than the 3D bulk (such as 1D hinge state or 0D corner state), has unveiled a novel unconventional bulk-boundary correspondence that extends beyond the conventional one.
Recently, higher-order topological Weyl semimetals have been  proposed~\cite{PhysRevLett.125.146401,PhysRevLett.125.266804,RN320}, where the conventional and second-order WPs are respectively defined as critical points that  separate  quantum anomalous Hall insulator (QAHI) and normal insulator and that separate QAHI and a higher-order quadrupole topological insulator (HOQTI) in momentum space~\cite{doi:10.1126/science.aah6442,PhysRevLett.125.266804}.
This definition further implies  that a combination of conventional and second-order  WPs can give rise to a nontrivial Weyl dipole~\cite{PhysRevLett.125.266804,PhysRevLett.124.104301}, which separates a normal insulator and a HOQTI. Such a nontrivial  Weyl dipole is characterized by a second-order $Z_2$ topological charge, namely, a quantized quadrupole moment $Q_{WD}$, as illustration in Fig.~\ref{fig1}(a).
Therefore, the $Z_2$ Weyl dipole should also come in pairs, similar to the WPs and the $Z_2$ nodal lines.
Moreover, the  $Z_2$ topological charge guarantees the appearance of second-order boundary states along the hinge of the TWDPs, as depicted in Fig.~\ref{fig1}(b).
In Ref.~\cite{PhysRevB.109.L081101}, the Weyl dipole protected by a staggered Chern number is also proposed and is termed as dipolar Weyl point.
However, it is important to note that while  both higher-order topological Weyl  semimetals  and the TWDPs exhibit hinge states, most of the higher-order topological Weyl  semimetals lack nontrivial Weyl dipoles, and thus can not be classified as TWDPs ~\cite{PhysRevLett.125.146401,RN320,RN319,PhysRevB.109.035148}. Despite the realization of various higher-order topological Weyl  phases in the artificial crystals and phonon spectrum~\cite{PhysRevLett.125.266804,RN319,RN320,RN322}, a TWDP protected by a quantized quadrupole moment has yet to be reported in a crystalline material.

Drawing an analogy with electronic systems, the concept of topology has been applied to bosonic particles~\cite{https://doi.org/10.1002/adfm.201904784,doi:10.1126/sciadv.abd1618,RN323,RN324,https://doi.org/10.1002/adfm.202401684}. Thus far, various topological phonon band crossings have been proposed, including phonons with nodal points~\cite{PhysRevLett.121.035302,PhysRevLett.120.016401,PhysRevLett.123.065501,PhysRevLett.122.104302,PhysRevLett.124.105303,PhysRevLett.126.185301,PhysRevLett.131.116602,
https://doi.org/10.1002/advs.202207508,RN216} and high-dimensional geometric shapes~\cite{PhysRevLett.123.245302,RN108,RN106,RN318,10.1063/5.0095281,doi:10.1126/science.adf8458}. Over 10,000 crystalline materials have been predicted to be topological phononic materials from the theory~\cite{doi:10.1126/science.adf8458}, and some of them have been verified from the experiment~\cite{RN216,PhysRevLett.123.245302,RN108,RN106}.
It is feasible to integrate higher-order topology into topological phonons, such as Weyl phonons, and the exploration of higher-order topological phonons, specifically higher-order Weyl phonons, has emerged as a prominent focus in condensed-matter research~\cite{PhysRevLett.125.146401,RN325,PhysRevLett.126.156801,https://doi.org/10.1002/adma.202407437,PhysRevLett.133.176602,PhysRevB.109.035148}.


In this work, we focus on the TWDP in the phonon spectrum of the crystalline materials and discover an unconventional TWDP in Y(OH)$_3$ with a space group (SG) $P6_3$ (No. 173).
The unconventional  TWDP is distinguished from all previously reported higher-order Weyl phases~\cite{PhysRevLett.125.146401,PhysRevLett.125.266804,RN319,RN320,RN322,PhysRevLett.130.116103}.
The two Weyl dipoles (WDs) in  Y(OH)$_3$  are connected by time-reversal symmetry (${\cal T}$), and each WD   is formed by  a charge-3 (C-3) WP with ${\cal{C}}=-3$ on the  $\Gamma$-$A$ path, along with three  charge-1 (C-1) WPs with ${\cal{C}}=1$ on the three  $M$-$L$ paths. The $k_z$ plane, which is present both inside and outside of the two WDs, exhibits a nontrivial quantized quadrupole moment of $1/2$ ($q_{xy}$ = $1/2$) and a trivial quadrupole moment of 0 ($q_{xy}$ = $0$), respectively. This directly shows that the WD in   Y(OH)$_3$ is a nontrivial $Z_2$ WD. Due to the presence of both first- and second-order topological charges, the boundary states in crystal Y(OH)$_3$ are also quite unique. Y(OH)$_3$ features 2D sextuple-helicoid Fermi-arc states on the top and bottom surfaces as well as 1D hinge states that connect the two WDs along the side hinges.

\begin{figure}
\includegraphics[width=8.5cm]{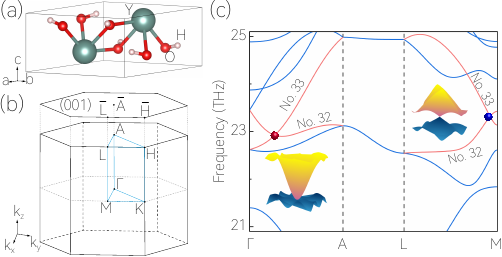}
\caption{(a) Crystal structure of Y(OH)$_3$. (b) Bulk BZ and (001) surface BZ of Y(OH)$_3$. (c) Enlarged phonon dispersion curve of Y(OH)$_3$ in the frequency range of 21-25 THz. In (c), the red and blue balls represent the C-3 WP and C-1 WP, respectively. The 3D plots of the phonon dispersion curve around the two doubly degenerate crossing points are also exhibited.
\label{fig2}}
\end{figure}

\textcolor{blue}{\textit{Unconventional WP and WD phonon in Y(OH)$_3$.--}}The crystal structure of crystalline material Y(OH)$_3$, which belongs to SG No. 173, is of the hexagonal-type, as shown in Fig.~\ref{fig2}(a).
The operators in SG No. 173 are generated by a six-fold screw rotation $\{C_{6z}|00\frac{1}{2}\}$, which results in the emergence of $C_{3z}$ and $\{C_{2z}|00\frac{1}{2}\}$.
Moreover, the Y(OH)$_3$ has ${\cal T}$.
The  optimized lattice constants of Y(OH)$_3$ are $a = b = 6.19$ \AA~and $c = 3.50$ \AA. The Wyckoff positions of Y, O, and H atoms are $2b$, $6c$, and $6c$, respectively. The bulk Brillouin zone (BZ) and (001) surface BZ of Y(OH)$_3$ are illustrated in Fig.~\ref{fig2}(b).
The Y(OH)$_3$ is  dynamically stable, as indicated by its phonon spectrum, which does not exhibit any imaginary frequency modes, as shown in Fig. S1 in the Supplementary Material (SM)~\cite{supp}. Information regarding the computational methods is also  available in SM~\cite{supp}.

\begin{figure}
\includegraphics[width=8.5cm]{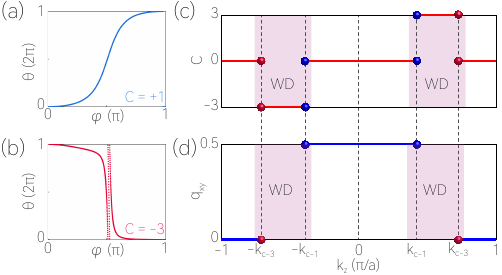}
\caption{(a) and (b) Evolutions of the average position of the Wannier center for C-1 and C-3 WPs. (c) and (d) Evolutions of the Chern number $\cal{C}$ and quadrupole moment $q_{xy}$ for the gaps between Nos. 32 and 33 phonon bands along the $k_z$.
\label{fig3}}
\end{figure}

Here, we concentrated on the phonon bands within the frequency range of $21$ to $25$ THz, as illustrated in Fig.~\ref{fig2}(c). Within this range, we observed  two doubly degenerate WPs located on the $\Gamma$-$A$ and $L$-$M$ paths, formed by 32nd and 33rd bands, which are represented by red and blue balls, respectively.
A careful scan reveals that apart from these two kinds of WPs, the 32nd and 33rd bands did not form any other band degeneracies.
Thus, there are only eight WPs within the target frequency range: two WPs on the $\Gamma$-$A$ path connected by ${\cal T}$, and six WPs on the $L$-$M$ paths connected by the $C_{3z}$ and ${\cal T}$, as illustrated in Fig.~\ref{fig1}(b).
Then, according to the Nielsen-Ninomiya no-go theory~\cite{NIELSEN198120,NIELSEN1981173} and the encyclopedia of emergent particles~\cite{YU2022375}, we can conculde that the WPs on the $\Gamma$-$A$ path are unconventional C-3 WPs with the same Chern number, as the WPs on the $L$-$M$ path must be C-1 WPs~\cite{YU2022375}.
This is confirmed by the Wilson-loop  calculations, which is performed on the spheres enclosing the WPs.
As shown in Figs.~\ref{fig3}(a) and~\ref{fig3}(b), the WP  on the $\Gamma$-$A$ path possesses a Chern number of ${\cal{C}}=-3$, whereas the WPs on the $L$-$M$ path have ${\cal{C}}=1$.
Moreover, we find that the $k_z$-positions of the C-3 WPs and C-1 WPs are located at $k_{\rm{C-3~WP}}\simeq\pm0.272\pi$ and $k_{\rm{C-1~WP}}\simeq\pm 0.08\pi$, respectively.


A 3D system can be viewed as a collection of infinite 2D subsystems that depend on $k_z$.
Generally, one can calculate the Wilson loop and nested Wilson loop to identify the first-order and second-order topological properties of these 2D subsystems, respectively~\cite{PhysRevLett.125.146401,PhysRevLett.125.266804}.
We first study the first-order topology of the  $k_z$-fixed 2D subsystems by calculating the Chern number.
The Chern number as a function of $k_z$ is plotted in Fig.~\ref{fig3}(c).
One observes that   in the regions where $|k_z|>|k_{\rm{C-3~WP}}|$ and $|k_z|<|k_{\rm{C-1~WP}}|$, the 2D planes  have  ${\cal{C}}=0$. However, in the region where $|k_{\rm{C-1~WP}}|<|k_z|<|k_{\rm{C-3~WP}}|$, the planes have Chern numbers of  ${\cal{C}}=\pm3$.

For the 2D subsystems with ${\cal{C}}=0$, they have  trivial first-order topological properties, but they may possess nontrivial higher-order topological properties.
Particularly, the combination of  one C-3 WP and three C-1 WPs above (below) $k_z=0$ plane can be considered as a unconventional WD.
Since the Chern number of the  WD is zero, the WD has a trivial first-order topology. However, due to the presence of $C_{2z}$,  each $k_z$ plane is featured with a second-order topological charge known as quantized quadrupole moment $q_{xy}(k_z)$. $q_{xy}(k_z)$ has two topologically inequivalent values: 0 and $1/2$.
$q_{xy}=0$ and $q_{xy}=1/2$ correspond to a normal insulator and a HOQTI, respectively.
Therefore, the topological character of the WD can be characterized by the difference of the  quantized quadrupole moment on two $k_z$ planes that enclose the WD,
\begin{equation}
    Q_{WD}=q_{xy}(k_z^>)-q_{xy}(k_z^<) \ \ \ {\rm mod} \ \ \ 1,
\end{equation}
where  $|k_z^{>}|\in (0.272\pi, \pi)$, $|k_z^{<}|\in (0, 0.08\pi)$, and  $q_{xy}(k_z)$ is encoded in the determinant of the nested Wilson loop operator obtained in each $k_z$ plane~\cite{doi:10.1126/science.aah6442}.

As shown in Fig.~\ref{fig3}(d), one finds that the planes in the region where $|k_z|<|k_{\rm{C-1~WP}}|$ host a nontrivial $q_{xy}=1/2$, whereas the planes in the region where $|k_z|>|k_{\rm{C-3~WP}}|$ host a trivial $q_{xy}=0$.
Therefore, the topological charges of the two unconventional WDs are obtained as $Q_{WD}=1/2$, showing the two unconventional WDs are  topologically nontrivial.
This also implies that the WPs in Y(OH)$_3$ are more robust than those in conventional Weyl semimetals.
It is well known that two WPs with opposite Chern numbers will annihilate when they come together.
However, in stark contrast, even if the C-3 WP and the three C-1 WPs above (below) $k_z=0$ plane meet together, they will not annihilate to form an insulator phase, as they as a whole remain nontrivial and are topologically protected by the quantized quadrupole moment.

Hence, the Y(OH)$_3$ exhibits a variety of topological phases. The planes in the region where $|k_z|>|k_{\rm{C-3~WP}}|$, with $\cal{C}$ = 0 and $q_{xy}$ = 0, are NIs. The planes in the region where $|k_{\rm{C-1~WP}}|<|k_z|<|k_{\rm{C-3~WP}}|$, with $\cal{C}$ = $\pm3$, are QAHIs~\cite{PhysRevLett.61.2015,doi:10.1126/science.1234414,RevModPhys.95.011002}, which belong to first-order 2D topological insulators. The planes in the region where $|k_z|<|k_{\rm{C-1~WP}}|$, with $\cal{C}$ = 0 and $q_{xy} = 1/2$, are HOQTIs~\cite{doi:10.1126/science.aah6442,PhysRevB.96.245115}, which belong to second-order 2D topological insulators.
At the boundaries of these insulator phases are the C-3 and C-1 WPs, which give rise to unconventional WDs.

\begin{figure}
\includegraphics[width=8.5cm]{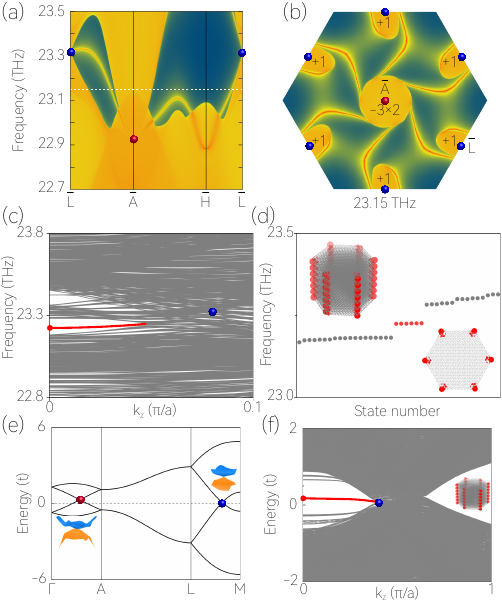}
\caption{(a) Projected spectrum of Y(OH)$_3$ on the (001) surface. (b) Constant-frequency slice at 23.15 THz. (c) Enlarged phonon spectrum (in the $k_z$ direction) within the frequency ranges of 22.8-23.8 THz for a sample of Y(OH)$_3$ with a tube-like geometry. The hinge states are highlighted by the red curves. (d) Frequency spectrum for a sixfold degenerate state (red dot) at $k_z = 0$. These insets are the spatial distributions for the sixfold degenerate state. (e) Bulk band structure of the tight-binding model. These insets are 3D plots of bands around the C-3 WP and C-1 WP. (f) The spectrum of a 1D tube geometry (in the $k_z$ direction). The inset is the spatial distribution for the state at the $k_z$ = 0.
\label{fig4}}
\end{figure}

\textcolor{blue}{\textit{Unconventional boundary states in Y(OH)$_3$.--}}
The coexistence of the first-order and second-order topologies is manifested in the multi-dimensional boundary modes. First, we calculated the surface spectrum of Y(OH)$_3$, which is projected on the (001) plane along the $\overline{L}$-$\overline{A}$-$\overline{H}$-$\overline{L}$ paths (see Fig.~\ref{fig4}(a)). As shown in Fig.~\ref{fig4}(b), the two C-3 WPs are projected to the same position, i.e., $\overline{A}$ surface point in the (001) surface BZ. Consequently, the two C-3 WPs necessitate the existence of six surface arcs connected to the  surface $\overline{A}$  point, resulting in 2D sextuple-helicoid arc-shaped surface states.

Second, we established a sample of Y(OH)$_3$ with a tube-like geometry using the phononic tight-binding model and calculated the phonon dispersion curve along the $k_z$ direction, as shown in Fig.~\ref{fig4}(c).
Since the two unconventional WDs have $Q_{WD}=1/2$, they will feature a second-order boundary state along the hinge of the Y(OH)$_3$ with a tube-like geometry.
Indeed, a phononic hinge band, shown by a red line, can be seen in Fig.~\ref{fig4}(c) and connects the  projections of unconventional WDs. To further confirm it, we select a sixfold degenerate state at the $k_z = 0$ (red dot in Fig.~\ref{fig4}(c)) and plot the spatial distribution for the state in Fig.~\ref{fig4}(d) under different viewpoints. Evidently, the sixfold degenerate state is distributed at the six 1D hinges of the tube sample, demonstrating the existence of the second-order boundary  states.

\textcolor{blue}{\textit{Discussions and conclusions.--}}
For a better understanding of the topological properties, we also develop a simple effective lattice model of the unconventional TWDP.
Based on the Y(OH)$_3$ material, we construct  a four-band tight-binding model for SG No. 173, in which two C-3 WPs and three C-1 WPs appear on the $\Gamma$-$A$ path and $L$-$M$ path, as shown in Fig.~\ref{fig4}(e). One C-3 WP and three C-1 WPs together form an unconventional WD with $Q_{WD}=1/2$,  which leads to the appearance of the hinge states [see Fig.~\ref{fig4}(f)].
The details of the Hamiltonian are given in SM~\cite{supp}.

In summary, with the help of first-principle calculation and symmetry analysis, we propose a new higher-order topological phonons, which integrates Weyl complex  phonon (a C-3 WP and three C-1 WPs), i.e., WD, with higher-order topology. $P6_3$ type Y(OH)$_3$ has been selected as the first crystalline material to host the TWDP in its optical phonon spectrum. Moreover, the two WDs in Y(OH)$_3$ exhibit the sextuple-helicoid surface and hinge boundary states from the nonzero Chern number ($|{\cal{C}}| = 6$) and the nontrivial quadrupole moment ($Q_{WD} = 1/2$), respectively. Our findings reveal a new type of higher-order topological phonons and demonstrate that the phonon spectra can serve as an excellent platform for investigating unconventional WDs and unconventional boundary states.

${Acknowledgements}$---J.W. and Y.W. contributed equally to this work. X.W. thanks Australian Research Council Discovery Early Career Researcher Award (Grant No. DE240100627) for support.

\bibliography{2.24}

\end{document}